\documentclass[sigconf,natbib=false]{acmart}
\pdfoutput=1
\usepackage{booktabs} 
\usepackage{totcount}
\usepackage{enumitem}
\usepackage{url}
\usepackage[style=numeric,giveninits=true,doi=false,isbn=false,url=false,maxnames=9]{biblatex}
\setlist{leftmargin=*}
\usepackage{xcolor}
\usepackage{quoting}

\setcopyright{cidr19}

\newcommand{\cameraready}[1]{{#1}}

\acmConference[CIDR'19]{Conference on Innovative Data Systems Research}{January 2019}{Asilomar, CA, USA}
\acmYear{2019}
\copyrightyear{2019}
\acmISBN{}
\acmDOI{}

\begin{document}
\sloppy

\title{Serverless Computing: One Step Forward, Two Steps Back}

\author{Joseph M. Hellerstein, Jose Faleiro, Joseph E. Gonzalez, Johann Schleier-Smith, Vikram Sreekanti, Alexey Tumanov and Chenggang Wu}
\affiliation{UC Berkeley}
\email{{hellerstein,jmfaleiro,jegonzal,jssmith,vikrams,atumanov,cgwu}@berkeley.edu}

\begin{abstract}
Serverless computing offers the potential to program the cloud in an autoscaling, pay-as-you go manner.
In this paper we address critical gaps in first-generation serverless computing, which place its autoscaling potential at odds with dominant trends in modern computing: notably data-centric and distributed computing, but also open source and custom hardware. 
Put together, these gaps make current serverless offerings a bad fit for cloud innovation and particularly bad for data systems innovation. 
In addition to pinpointing some of the main shortfalls of current serverless architectures, we raise a set of challenges we believe must be met to unlock the radical potential that the cloud---with its exabytes of storage and millions of cores---should offer to innovative developers.
\end{abstract}


\newtoggle{shownotes}
\toggletrue{shownotes}
\newtotcounter{numnotes}
\newcommand{\note}[3]{
  \iftoggle{shownotes}{\stepcounter{numnotes}\textcolor{#1}{#2: #3}}{}
}

\newcommand{\cuttext}[1]{}
\newcommand{\vikram}[1]{\note{olive}{vikram}{#1}}
\newcommand{\alexey}[1]{\note{brown}{alexey}{#1}}
\newcommand{\jmh}[1]{\note{red}{jmh}{#1}}
\newcommand{\chenggang}[1]{\note{purple}{chenggang}{#1}}
\newcommand{\jf}[1]{\note{orange}{jf}{#1}}
\newcommand{\joey}[1]{\note{cyan}{joey}{#1}}
\newcommand{\jss}[1]{\note{green}{jss}{#1}}

\newtoggle{showtodos}
\toggletrue{showtodos}
\newtotcounter{numtodos}
\newcommand{\todo}[1]{
  \iftoggle{showtodos}{
    \stepcounter{numtodos}\textcolor{blue}{\textbf{TODO: #1}}
  }{}
}

\newcommand{\checkforcrud}{
  \ifboolexpr{test{\ifnumcomp{\totvalue{numnotes}}{>}{0}} or
              test{\ifnumcomp{\totvalue{numtodos}}{>}{0}}}{
    \pagecolor{red!20}
  }{}
}

\newcommand{\smallitem}[1]{\vspace{0.3em}\noindent\textbf{#1}}
\newcommand{\smallitembot}{\vspace{0.5em}\noindent}

\newcommand{\imgdir}{./img}
\newcommand{\texdir}{./tex}

\maketitle

\renewcommand{\shortauthors}{Hellerstein et al.}


\section{Introduction}

Amazon Web Services recently celebrated its 12th anniversary, marking over a decade of public cloud availability. 
While the cloud began as a place to timeshare machines, it was clear from the beginning that it presented a radical new computing platform: the biggest assemblage of data capacity and distributed computing power ever available to the general public, managed as a service.

Despite that potential, we have yet to harness cloud resources in radical ways. 
The cloud today is largely used as an outsourcing platform for standard enterprise data services. 
For this to change, creative developers need programming frameworks that enable them to leverage the cloud's power. 

New computing platforms have typically fostered innovation in programming languages and environments. 
Yet a decade later, it is difficult to identify the new programming environments for the cloud.
And whether cause or effect, the results are clearly visible in practice: the majority of cloud services are simply multi-tenant, easier-to-administer clones of legacy enterprise data services like object storage, databases, queueing systems, and web/app servers. 
Multitenancy and administrative simplicity are admirable and desirable goals, and some of the new services have interesting internals in their own right. 
But this is, at best, only a hint of the potential offered by millions of cores and exabytes of data. 

%

\begin{figure}[t]
  \centering
    \includegraphics[width=0.4\textwidth]{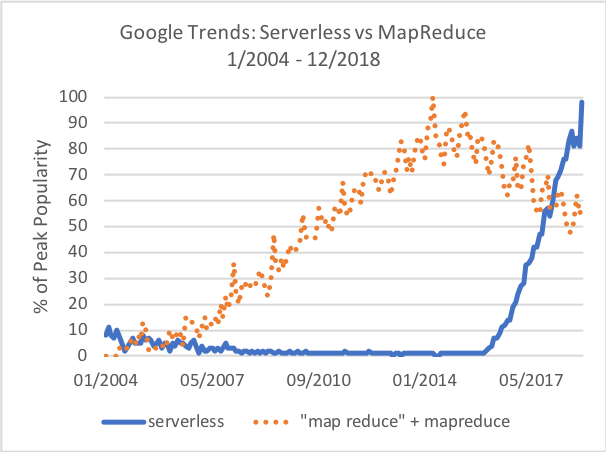}
  \caption{Google Trends for ``Serverless'' and ``Map Reduce'' from 2004 to time of publication.}
  \label{fig:trends}
\end{figure}

Recently, public cloud vendors have begun offering new programming interfaces under the banner of \emph{serverless computing}, 
and interest is growing. 
Google search trends show that queries for the term ``serverless'' recently matched the historic peak of popularity of the phrase ``Map Reduce'' or ``MapReduce'' (Figure~\ref{fig:trends}). 
There has also been a significant uptick in attention to the topic more recently from the research community~\cite{pywren, excamera, peeking-atc18, ephemeral-atc18}.
Serverless computing offers the attractive notion of a platform in the cloud where developers simply upload their code, and the platform executes it on their behalf as needed at any scale. 
Developers need not concern themselves with provisioning or operating servers, and they pay only for the compute resources used when their code is invoked.


The notion of serverless computing is vague enough to allow optimists to project any number of possible broad interpretations on what it might mean.
Our goal here is not to quibble about the terminology.
Concretely, each of the cloud vendors has already launched serverless computing infrastructure and is spending a significant marketing budget promoting it.
In this paper, we assess the field based on the serverless computing services that vendors are actually offering today and see why they are a disappointment as big as the cloud's potential.

\subsection{``Serverless'' goes FaaS}

To begin, we provide a quick introduction to \emph{Functions-as-a-Service} (FaaS), the commonly used and more descriptive name for the core of serverless offerings from the public cloud providers. 
Because AWS was the first public cloud---and remains the largest---we focus our discussion on the AWS FaaS framework, Lambda; offerings from Azure and GCP differ in detail but not in spirit.

The idea behind FaaS is simple and straight out of a programming textbook. 
Traditional programming is based on writing functions, which are mappings from inputs to outputs. 
Programs consist of compositions of these functions. 
Hence, one simple way to program the cloud is to allow developers to register functions in the cloud, and compose those functions into programs. 

Typical FaaS offerings today support a variety of languages (e.g., Python, Java, Javascript, Go), allow programmers to register functions with the cloud provider, and enable users to declare events that trigger each function.
The FaaS infrastructure monitors the triggering events, allocates a runtime for the function, executes it, and persists the results.
The user is billed only for the computing resources used during function invocation.

A FaaS offering by itself is of little value, since each function execution is isolated and ephemeral. 
Building applications on FaaS requires data management in both persistent and temporary storage, in addition to mechanisms to trigger and scale function execution. 
As a result, cloud providers are quick to emphasize that serverless is not only FaaS. 
It is FaaS supported by a ``standard library'': the various multitenanted, autoscaling services provided by the vendor\footnote{This might be better termed a ``proprietary library'', but the analogy to C's stdlib is apropos: not officially part of the programming model, but integral in practice.}. 
In the case of AWS, this includes S3 (large object storage), DynamoDB (key-value storage), SQS (queuing services), SNS (notification services), and more. 
This entire infrastructure is managed and operated by AWS; developers simply register FaaS code that uses these services and receive ``pay-as-you-go'' bills that scale up and down according to their storage and compute usage.

\subsection{Forward, but also Backward}

We emphasize that serverless computing provides a programming model that is not simply elastic, in the sense that humans or scripts can add and remove resources as needed; it is \emph{autoscaling}. 
The workload automatically drives the allocation and deallocation of resources. 
As modern applications increase in dynamics and complexity, the task of allocating VMs dynamically, monitoring services, and responding to workload changes becomes increasingly onerous, requiring constant human observation or bespoke scripts developed for individual applications.
By providing autoscaling, today's FaaS offerings take a big step forward for cloud programming, offering a practically manageable, seemingly unlimited compute platform\footnote{In many blog posts ostensibly about serverless computing, FaaS is combined with ``non-serverless'' services: i.e., services that do not autoscale, like AWS Elasticache.
Design patterns that use non-serverless services are out of scope of our discussion; one might even argue they are anti-patterns for serverless development since they do not autoscale.}.


Unfortunately, as we will see, today's FaaS offerings also slide two major steps backward. 
First, they painfully ignore the importance of efficient data processing. 
Second, they stymie the development of distributed systems. This is curious since data-driven, distributed computing is at the heart of most innovation in modern computing.
In the rest of the paper, we highlight the simple cases where FaaS currently offers some benefits.
We then elaborate on the shortcomings of the existing FaaS platforms alluded to above, and present straightforward use cases for which FaaS is incapable of providing an efficient way to get things done. 
Finally, we outline challenges in moving forward towards a fully-realized cloud programming infrastructure.


\section{Serverless is More? The easy cases}

AWS Lambda has been adopted by several applications looking to simplify their cloud deployments. 
Many of these use cases have been documented by Amazon~\cite{aws-lambda-studies}. 
This section provides an overview of the design patterns employed by these documented applications.

The use cases listed can be broadly divided into three categories based on the nature of interaction between function invocations.


\noindent \textbf{Embarrassingly parallel functions}. In some applications, each function invocation is an independent task and never needs to communicate with other functions. 
These uses of Lambda functions are inherently limited in their scope and complexity. 
Concrete examples include functions that resize the canonical versions of images for use on a variety of end-user devices (Seattle Times), perform object recognition in images (V!Studios), and run integer programming-based optimizations (Financial Engines)~\cite{aws-lambda-studies}. 
The PyWren~\cite{pywren} and ExCamera~\cite{excamera} research projects have shown that AWS Lambda can be made (with some effort) to perform a wider variety of such ``map'' functions, including some simple featurization and linear algebra workloads. 
Such applications can directly exploit Lambda's auto-scaling features to scale up or down on demand because independent requests never need to communicate with each other and require only small granules of computation.

\noindent \textbf{Orchestration functions}. A second class of use cases leverages serverless functions simply to orchestrate calls to proprietary autoscaling services, such as large-scale analytics. 
For instance, Amazon provides a reference application architecture that uses Lambda functions to orchestrate analytics queries that are executed by AWS Athena, an autoscaling query service that works with data in S3 \cite{athena-ex-1}. 
Another example uses Lambda functions to preprocess event streams before funneling them to Athena via S3 \cite{athena-ex-2}. 
In both applications the ``heavy lifting'' of the computation over data is done by Athena, not by Lambda functions. 
This enables efficient data manipulation at scale, by pushing computation into an existing autoscaling service.

Our experience developing Google Cloud Dataprep by Trifacta is exemplary of this design pattern as well~\cite{trifacta-debull18}. The basic architecture of Dataprep involves (1) web client software synthesizing programs in the domain-specific language (DSL) of Google Cloud Dataflow, (2) stateless autoscaling services in the Google Cloud that handle client requests and pass the DSL statements along to Dataflow, and (3) the autoscaling Dataflow service executing the DSL at scale.
Although the middle tier here is not implemented over Google's FaaS offering, similar architectures could use FaaS.

\noindent \textbf{Function Composition}. The third category consists of collections of functions that are composed to build applications and thus need to pass along outputs and inputs. 
Examples of such applications include workflows of functions chained together via data dependencies. 
These functions generally manipulate state in some way, and are designed as event-driven workflows of Lambda functions, stitched together via queueing systems (such as SQS) or object stores (such as S3). 
For instance, Autodesk's account creation platform makes several function invocations in the critical path of creating a single user account \cite{aws-lambda-studies}.
Each invocation handles a small portion of the account creation logic, such as email notification and validation. 
The authors of that case study reported average end-to-end sign-up times of \emph{ten minutes}; as we will see in the next section, the overheads of Lambda task handling and state management explain some of this latency. 

\smallitembot

In short, current FaaS solutions are attractive for simple workloads of independent tasks---be they embarrassingly parallel tasks embedded in Lambda functions, or jobs to be run by the proprietary cloud services. 
Use cases that involve stateful tasks have surprisingly high latency: 10 minutes is a long turnaround time for a cloud vendor to publicize, even for a sign-up workflow. 
These realities limit the attractive use cases for FaaS today, discouraging new third-party programs that go beyond the proprietary service offerings from the vendors. 


\cuttext{combining existing commodity services with smart clients. Our experience with Google Cloud Dataprep by Trifacta is exemplary of this design pattern~\cite{trifacta-debull18}. (Dataprep was not in fact implemented over Google's FaaS offering, but the core design is a serverless architecture.) Dataprep begins by serving a feature-rich ``fat client'' WebAssembly application to a user's web browser. The client requests modest amounts of data from standard cloud services (fileservers or databases) for visualization, and based on user interactions synthesizes programs that can be pushed to a commodity cloud service---Google Cloud Dataflow---for execution. The client calls are independent, stateless requests for code, data, or Dataflow execution. The Dataprep backend tasks require minimal cloud compute cycles and administration, and cost very little to run. Yet Dataprep is routinely used to author complex wrangling programs that run on multi-terabyte datasets and consume hours of vCPU time. The key here is that the ``heavy lifting'' of manipulating data at scale is done by the commodity Dataflow service. The innovation in Dataprep is in fact not on the cloud side; it is in the client's novel use of visualization, machine learning and code synthesis to leverage the existing cloud services.}

\cuttext{
These examples highlight the pros and cons of current FaaS offerings.  \vikram{This paragraph is a little hard to follow. I didn't feel like I came away from it with a clear sense of what the tangible cons actually were. The second half in particular felt pretty muddled.}
FaaS works best when each function invocation can be kept independent of others; such applications are a perfect fit for a stateless function invocation model.
On the other hand, stateful applications that require communication across Lambda invocations must be highly latency-tolerant due to overheads in the infrastructure coordinating Lambda invocations. 
Such applications are forced to make major tradeoffs of performance for ease of administration. 
For instance, although end-to-end request processing times on the order of minutes might be acceptable for Autodesk's user account creation (although it is an unusual delay for that functionality today), they would be unacceptable in any moderately interactive application. 
Similarly, the storage overheads associated with these applications may be unacceptable for applications that require significant throughput such as model training. 
\jmh{I think I disagree with the below in some ways. If you can synthesize code that works in the existing services, it does give you the ability to ship code to data, as in Trifacta. But it's limited and proprietary. Will work on it later.}
\vikram{You can ship code to data but not for general purpose computation. In some sense, I think this is why existing FaaS (and also Ray?) is an attractive paradigm: You can write arbitrary code and have it easily parallelized. Dataflow is good if you want to do MapReduce-y things. Athena is good if you have data that you can query with SQL---what if you don't? }
The orchestration use cases are in some sense indicative of the severity of these limitations; developers are restricted to writing no more than ``Scripts-as-a-service'' to cobble together an application out of proprietary ``heavy lifting'' components. 
Furthermore, this pattern is symptomatic of a slippery slope of vendor lock-in; developers have no choice but to rely on a cloud provider's specialized high-performance systems because such systems are impossible to replace given the building blocks on hand. }


\section{Why Serverless Today is Too Less}

The cloud offers three key features: unlimited data storage, unlimited distributed computing power, and the ability to harness these only as needed---paying only for the resources you consume, rather than buying the resources you might need at peak.

Serverless computing takes one step forward and two steps back from this vision.         
It realizes the potential of pay-as-you-go, fully-managed execution of end-user code via autoscaling.
Unfortunately, as we will see in this section, current FaaS offerings fatally restrict the ability to work efficiently with data or distributed computing resources. 
As a result, serverless computing today is at best a simple and powerful way to run embarrassingly parallel computations or harness proprietary services.
At worst, it can be viewed as a cynical effort to lock users into those services and lock out innovation. 


The list of limitations in today's FaaS offerings is remarkable.
Our running example, AWS Lambda, has the following constraints that are typical of the other vendors as well\footnote{\cameraready{Our discussion in this paper represents the state of AWS Lambda as of Fall 2018. Some new details were announced at the AWS re:Invent conference in late November 2018. We have only had time before publication deadline to comment on them briefly where they may affect our reported results. These announcements do not seem to address our main concerns in a substantive way.}}:
\begin{enumerate}
    \item \textbf{Limited Lifetimes.} After \cameraready{15} minutes, function invocations are shut down by the Lambda infrastructure. 
    Lambda may keep the function's state cached in the hosting VM to support ``warm start'', but there is no way to ensure that subsequent invocations are run on the same VM. 
    Hence functions must be written assuming that state will not be recoverable across invocations.
    
    \item \textbf{I/O Bottlenecks.} Lambdas connect to cloud services---notably, shared storage---across a network interface. 
    In practice, this typically means moving data across nodes or racks.
    With FaaS, things appear even worse than the network topology would suggest.
    Recent studies show that a single Lambda function can achieve on average 538Mbps network bandwidth; numbers from Google and Azure were in the same ballpark~\cite{peeking-atc18}. 
    This is an order of magnitude slower than a single modern SSD.
    Worse, AWS appears to attempt to pack Lambda functions from the same user together on a single VM, so the limited bandwidth is shared by multiple functions. 
    The result is that as compute power scales up, per-function bandwidth shrinks proportionately. 
    With 20 Lambda functions, average network bandwidth was 28.7Mbps---2.5 orders of magnitude slower than a single SSD~\cite{peeking-atc18}\footnote{\cameraready{AWS announced the availability of 100Gbps networking on 11/26/2018. This moves the needle but leaves the problem unsolved: once you hit the cap, you are constrained. Even with 100Gbps/64 cores, under load you get ${\sim} 200$MBps per core, still an order of magnitude slower than a single SSD.}}. 
    
    \item \textbf{Communication Through Slow Storage.} While Lambda functions can initiate outbound network connections, they themselves are not directly network-addressable in any way while running. 
    As a result, two Lambda functions can only communicate through an autoscaling intermediary service; today, this means a storage system like S3 that is radically slower and more expensive than point-to-point networking. 
    As a corollary, a client of Lambda cannot address the particular function instance that handled the client's previous request: there is no ``stickiness'' for client connections. 
    Hence maintaining state across client calls requires writing the state out to slow storage, and reading it back on \emph{every} subsequent call.
    
    \item \textbf{No Specialized Hardware.} FaaS offerings today only allow users to provision a timeslice of a CPU hyperthread and some amount of RAM; in the case of AWS Lambda, one determines the other.
    There is no API or mechanism to access specialized hardware.
    However, as explained by Patterson and Hennessy in their recent Turing Lecture~\cite{turing-talk-patterson}, hardware specialization will only accelerate in the coming years.
    
\end{enumerate}

These constraints, combined with some significant shortcomings in the standard library of FaaS offerings, substantially limit the scope of feasible serverless applications.
A number of corollaries follow directly.

    \noindent \textbf{FaaS is a Data-Shipping Architecture.} This is perhaps the biggest architectural shortcoming of FaaS platforms and their APIs. 
    Serverless functions are run on isolated VMs, separate from data. 
    In addition, serverless functions are short-lived and non-addressable, so their capacity to cache state internally to service repeated requests is limited.
    Hence FaaS routinely ``ships data to code'' rather than ``shipping code to data.'' 
    This is a recurring architectural anti-pattern among system designers, which database aficionados seem to need to point out each generation. 
    Memory hierarchy realities---across various storage layers and network delays---make this a bad design decision for reasons of latency, bandwidth, and cost.
  
    \noindent \textbf{FaaS Stymies Distributed Computing.} Because there is no network addressability of serverless functions, two functions can work together serverlessly only by passing data through slow and expensive storage.
    This stymies basic distributed computing.
    That field is founded on protocols performing fine-grained communication between agents, including basics like leader election, membership, data consistency, and transaction commit. 
    Many distributed and parallel applications---especially in scientific computing---also rely on fine-grained communication.
    
    One might argue that FaaS encourages a new, event-driven distributed programming model based on global state. 
    But it is well-known that there is a duality between processes passing messages, and event-driven functions on shared data~\cite{lauer1979duality}. 
    For FaaS, event handling still requires passing pieces of the global state from slow storage into and out of stateless functions, incurring time and cost. 
    Meanwhile, the current serverless storage offerings offer weak consistency across replicas.
    Hence in the event-driven pattern, agreement across ephemeral functions would still need to be ``bolted on'' as a protocol of additional I/Os akin to classical consensus. 
    
    In short, with all communication transiting through storage, there is no real way for thousands (much less millions) of cores in the cloud to work together efficiently using current FaaS platforms other than via largely uncoordinated (embarrassing) parallelism.
    
   
    \noindent \textbf{FaaS stymies hardware-accelerated software innovation.}
    Many of the main Big Data use cases today leverage custom hardware. 
    The most prominent example is the pervasive use of GPUs in deep learning, but there is ongoing innovation in the use of accelerators for database processing as well. 
    Current FaaS offerings all run on a uniform and fairly mundane virtual machine platform.
    Not only do these VMs not offer custom processors, the current FaaS offerings do not even support main-memory databases that maintain large RAM-based data structures---the largest Lambda instance only allows for 3GB of RAM. 
    The lack of access to such hardware---along with appropriate pricing models---significantly limits the utility of FaaS offerings as a platform for software innovation.
    
    \noindent \textbf{FaaS discourages Open Source service innovation.} Most popular open source software cannot not run at scale in current serverless offerings. 
    Arguably this is inherent: that software was not designed for serverless execution, and expects human operation.
    But given the FaaS limitations on data processing and distributed computing, one should not expect new scalable open source software to emerge. 
    In particular, open-source data systems---an area of rapid growth and maturity in recent years---would be impossible to build on current FaaS offerings. 
    Current serverless infrastructure, intentionally or otherwise, locks users into either using proprietary provider services or maintaining their own servers.




\subsection{Case Studies}\label{sec:case-studies}
To evaluate the severity of these problems, we document three case studies from Big Data and distributed computing settings. 

\smallitem{Model Training}.
Our first case study explores AWS Lambda's performance for a common data-intensive application: machine learning model training.
As we will see, it suffers dramatically from the data-shipping architecture of Lambda.

Using public Amazon product review data~\cite{amazon-review}, we configured TensorFlow to train a neural network that predicts average customer ratings. 
Each product review is featurized with a bag-of-words model, resulting in 6,787 features and 90GB total of training data.
The model is a multi-layer perceptron with two hidden layers, each with 10 neurons and a Relu activation function.
Each Lambda is allocated the maximum lifetime (15 min) and 640MB RAM and runs as many training iterations as possible.
Our training program uses the \emph{AdamOptimizer} with a learning rate of 0.001 and a batch size of 100MB.

Each iteration in Lambda took 3.08 seconds: 2.49 to fetch a 100MB batch from S3 and 0.59 seconds to run the \emph{AdamOptimizer}.
A Lambda function times out of its 15-minute limit after 294 iterations of this algorithm.
We trained the model over 10 full passes of the training data, which translates to 31 sequential lambda executions, each of which runs for 15 minutes, or \emph{465 minutes total latency}. 
This costs \$0.29. 

For comparison, we trained the same model on an \texttt{m4.large} EC2 instance, which has 8GB of RAM and 2vCPUs.
In this setting, each iteration is significantly faster (0.14 seconds): 0.04 seconds to fetch data from an EBS volume and 0.1 seconds to run the optimizer.
The same training process takes about 1300 seconds (just under 22 minutes), which translates to a cost of \$0.04.
Lambda's limited resources and data-shipping architecture mean that running this algorithm on Lambda is 21$\times$ slower and 7.3$\times$ more expensive than running on EC2.

\begin{table*}
    \centering
    \begin{small}
    \begin{tabular}{| r | c | c | c | c | c |  c | c |}
        \hline
            &  Func. Invoc. &  Lambda I/O & Lambda I/O & EC2 I/O &  EC2 I/O &  EC2 NW\\
            &   (1KB)  & (S3)  & (DynamoDB) &  (S3) & (DynamoDB) & (0MQ)  \\\hline
        Latency & 303ms & 108ms & 11ms & 106ms & 11ms & 290$\mu$s \\
        Compared to best &  1,045$\times$ & 372$\times$ & 37.9$\times$ &  365$\times$  & 37.9$\times$ & 1$\times$ \\ \hline
    \end{tabular}
    \end{small}
    \captionsetup{textfont={rm,small}}
    \caption{\textbf{Latencies}. We compare the latency of ``communicating'' 1KB in various ways. To model pure functional event-driven communication, we show the cost of invoking a no-op Lambda function on a 1KB argument, averaged over 1,000 calls. We then show the cost of two explicit 1KB I/Os (write+read) from Python Lambda function and an EC2 instance to S3 and DynamoDB, averaged across 5k trials. Finally we show the cost of direct messaging by measuring a 1KB network message roundtrip, measured using python and the ZeroMQ message library running across two EC2 instances, averaged across 10k trials.
    \vspace{-2em}}
    \label{table:latency}.
\end{table*}

\smallitem{Low-Latency Prediction Serving via Batching.}
Our next case study focuses on the downstream use of a trained model: making live predictions.
We have been working for some time on low-latency prediction serving in Clipper~\cite{clipper}.
At first glance, prediction serving appears to be well-suited to FaaS. 
Each function is independent, and multiple copies can be deployed to scale the number of predictions made with a certain model. 

In practice, prediction serving relies on access to specialized hardware like GPUs, which are not available through AWS Lambda.
Setting that issue aside, we wanted to understand if the key performance optimizations of a system like Clipper could be achieved in a FaaS setting.
One optimization in Clipper is to process inputs in batches; in the traditional case this provides pipeline parallelism across the handling of input requests (performed by a CPU) with the multi-input vector processing of prediction (performed by a GPU). 
We were curious to see whether Lambda could provide similar benefits for pipelining batch accumulation with prediction.

To that end, we exercised Lambda's favored service for batching inputs: AWS Simple Queueing Service (SQS). 
We wrote a simple application on Lambda that pipelines batching work done by SQS with a trivial classifier running in Lambda functions. 
Specifically, our application accepts batches of text documents, uses a model to classify each word in the document as ``dirty'' or not, and writes the document out to a storage service with the dirty words replaced by punctuation marks. 
Our model in this experiment is a simple blacklist of dirty words.
SQS only allows batches of 10 messages at a time, so we limited all experiments here to 10-message batches.

The average latency over 1,000 batch invocations for the Lambda application was 559ms per batch if the model was retrieved on every invocation and results written back to S3.
As an optimization, we allowed the model to be compiled into the function itself and results were placed back into an SQS queue; in this implementation the average batch latency was 447ms.
For comparison, we ran two experiments using \texttt{m5.large} EC2 instances.
The first replaced Lambda's role in the application with an EC2 machine to receive SQS message batches---this showed a latency of 13ms per batch averaged over 1,000 batches---27$\times$ faster than our ``optimized'' Lambda implementation. 
The second experiment used ZeroMQ to replace SQS's role in the application, and receive messages directly on the EC2 machine. 
This ``serverful'' version had a per batch latency of 2.8ms---127$\times$ faster than the optimized Lambda implementation.

\cameraready{Pricing adds insult to the injury of performance in these services.}
If we wanted to scale this application to 1 million messages a second, the SQS request rate alone would cost \$1,584 per hour. 
Note that this does not account for the Lambda execution costs.
The EC2 instance on the other hand has a throughput of about 3,500 requests per second, so 1 million messages per second would require 290 EC2 instances, with a total cost of \$27.84 per hour---a 57$\times$ cost savings.

\smallitem{Distributed Computing}.
Lambda forbids direct network connectivity between functions, so we are forced to try alternative solutions to achieve distributed computation. 
As we will see, the available solutions are untenably slow.

As discussed in the previous section, there are two classical dual patterns to implement concurrent communicating systems: event-driven execution over shared state (the natural FaaS approach), or message-passing across long-running agents with distributed state~\cite{lauer1979duality}. 
In the Lambda environment, both design patterns share the property that functions can only pass data to each other through shared storage:  S3 or DynamoDB. 
In the event-driven pattern, data is read from and written to storage at the beginning and end of the function. 
In the message-passing pattern, messages are sent by writing to storage and read from storage via periodic polling.

We begin with a simple question: Is cloud storage a reasonable communication medium? 
To assess this, we measured ``send/receive'' (write+read) latency for communicating 1KB between Python functions. 
The Lambda results were inordinately slow, as shown in Table~\ref{table:latency}. 
They come in two forms. 
First we measure the pure functional, event-driven programming cost of a 1KB object being handled by a Lambda function invocation---this incurs both I/O and function overheads and is exorbitantly expensive\footnote{\cameraready{On 11/26/2018 AWS announced \emph{Firecracker}, a microVM framework that supports 125ms startup time for vanilla VMs. This would have at best modest effects on our results in Table~\ref{table:latency}; it is still orders of magnitude slower than traditional network messaging.}}.
Next we measure the cost of explicit I/O from Lambda, namely an average write+read from a long-running function---this is still over an order of magnitude slower than one would like.
We see that latencies from EC2 are almost identical, so the overhead is in the storage service costs, not in Lambda. 
Finally we show the latency of (acked) messaging using Python functions directly addressing each other via ZeroMQ. This last cost is close to typical intra-rack datacenter network measurements; studies from even a few years ago report average inter-rack measurements around 1.26ms~\cite{guo2015pingmesh}
In sum, communicating via cloud storage is not a reasonable replacement for directly-addressed networking, even with direct I/O---it is at leaste one order of magnitude too slow. 
``Pure'' functional FaaS programming style exacerbates that expense to an inordinate degree, and should be avoided at all costs.

To put this into perspective, we constructed a distributed systems case study in the style of communicating agents. 
As noted in the previous section, regardless of which design pattern you choose, distributed agreement on state (even ``global'' state in AWS' loosely-consistent storage) requires some kind of protocol. 
There is a broad literature of distributed protocols, but most require at least agreeing on a leader or the membership of the system at any time. 
To model this, we implemented one of the simplest of these protocols in Python: Garcia-Molina's bully leader election~\cite{garcia1982elections}. 
Using Lambda, all communication between our functions was done in blackboard fashion via DynamoDB.
With each Lambda polling four times a second, we found that \emph{each round of leader election took 16.7 seconds}.
Recall that Lambda functions are killed after \cameraready{15} minutes. 
Hence in the (unachievable) best-case scenario---when each leader is elected immediately after it joins the system---the system will spend \cameraready{1.9\%} of its aggregate time simply in the leader election protocol.
Even if you think this is tolerable, note that using DynamoDB as a fine-grained communication medium is incredibly expensive: Supporting a cluster of 1,000 nodes costs at minimum \$450 per hour\footnote{In ``election mode''---when leader election is happening---each node does about 10 reads every time it polls the storage medium, which happens 4 times a second. 
In non-election mode, each node does 2 reads per polling cycle. 
Our cost estimate represents the best case scenario, in which each leader is elected immediately after joining the cluster---in practice, costs might be much higher.}.

\subsection{Can Limitations Set Us Free?}

The limitations documented above render today's FaaS frameworks untenable for building sophisticated new backend functionality.
Still, limitations on developers are not necessarily a bad thing. 
Sometimes it is important for a new platform to reflect its ``physics'' in ways that encourage developers to write programs well-suited to the platform. 
In particular, FaaS limitations favor operational flexibility over developer control, a theme we generally agree is critical to the scale and elasticity of the cloud---and a major design shift from the traditional data systems ethos. 
Are some of the limitations of FaaS actually healthy for the future of distributed programming?

One benefit is that constraints can lead to deeper innovation.
As the most prominent example, today's FaaS frameworks offer few guarantees regarding sequential execution across functions. 
Developers are forced to compose larger programs out of asynchronous tasks, with no guarantees like sequential consistency or serializability to reason about the semantics of global state mutation across tasks.
These limitations can be challenging for developers used to writing sequential programs or transactions. 
But the result may be both healthy and manageable: this kind of ``disorderly'' loosely-consistent model has been at the heart of a number of more general-purpose proposals for scalable, available program design in recent years, including from our group~\cite{campbellhelland, bloom, shapiro2011conflict}.
As another example, today's FaaS frameworks offer no guarantees of physical hardware locality: developers cannot control where a function will run, or if its physical address will even remain constant. 
Again, this may be a manageable constraint: virtual addressing of dynamically shifting agents was a hallmark of prior work in the peer-to-peer research that presaged cloud services~\cite{ratnasamy2001scalable,stoica2002internet}. 

Another benefit of constraints is simplicity, which in turn can foster platform development and community. 
There is a constructive analogy here to MapReduce: while not a success in its own right, MapReduce changed the mindset of the developer community and eventually led to the reinvigoration of SQL and relational algebra (in the form of ``dataframe'' libraries) as popular, scalable interfaces for programming sophisticated analytics. 
Perhaps today's FaaS offerings will similarly lead to the reinvigoration of prior ideas for distributed programming at scale. 
A notable difference is that SQL and relational algebra were well established, whereas the natural end-state for asynchronous distributed programming over data in the cloud remains a matter of research and debate. 

Some of the constraints of current FaaS offerings can also be overcome, we believe, maintaining autoscaling while unlocking the performance potential of data and distributed computing.
In the next section we outline what we see as the key challenges and opportunities for moving forward on all three fronts.

\subsection{Early Objections}
\cameraready{
In this paper we purposely focused on the limitations of public FaaS APIs, and argued that they are disappointingly far from ready for general-purpose, data-rich programming. 
While many of our early readers concurred with the challenges we have raised, we have also heard objections. 
We try to address the most common ones here.
}

\vspace{1em}
\noindent
\emph{``You keep using that word. I do not think it means what you think it means.''}
\newline
\noindent
\cameraready{
Some of our colleagues working at the cloud platforms have argued that our view of the term ``serverless'' is too narrow:
behind the curtain of a cloud provider, they are building ``serverless'' solutions for customers that are autoscaling and management-free.
We understand that use of the term---we talk about Google Cloud Dataprep similarly. Perhaps this confusion is a reason why ``serverless'' is not a useful adjective for rallying the technical community around cloud innovation.
Simply put, \emph{the delivery of a particular special-purpose autoscaling backend service does not solve the problem of enabling general-purpose cloud programming}. 
Moreover, the work required to deliver these ``serverless'' backend offerings is done largely with old-fashioned ``node-at-a-time'' programming and is a traditional and expensive endeavor. 
Indeed, anyone can do this kind of work in the public cloud without ``serverless'' offerings, by using orchestration platforms like Kubernetes\footnote{By the same token, the peer-to-peer architectures of the turn of the century were also ``serverless''.}. 
Our goal here is to spark deeper discussion on the grand challenge of cloud-scale programming---one that we believe the public-facing FaaS offerings are trying (and presently failing) to provide.
}

\vspace{1em}
\noindent
\emph{``Just wait for the next network announcement!''}
\newline\noindent
\cameraready{Some readers of early drafts commented that datacenter networks are getting faster, and cloud providers are passing those innovations on to customers. 
Moreover, it is now conventional wisdom that scaling needs to be achieved by separating compute and data tiers.}

\cameraready{We have no argument with these points, but they do not address the key problems we raise here beyond matters of tuning. 
Datacenter networks will surely improve, yet inevitably will continue to play a limiting role in a larger memory hierarchy. 
Any reasonable system design will need the ability to selectively co-locate code and data on the same side of a network boundary, whether that is done via caching/prefetching data near computation or pushing computation closer to data. 
Neither feature is provided in a meaningful way by today's FaaS offerings. 
Improvements in network speeds are unlikely architectural game-changers; they can shift the parameters of using certain optimizations, but rarely justify the absence of those optimization opportunities. 
Meanwhile, \emph{separating compute and storage tiers in a logical design should not prevent co-location in a physical deployment}; one can scale compute and storage independently for flexibility \emph{and} colocate them as needed for performance. This is the heart of architectural indirections like ``data independence''---they increase flexibility rather than limit it.}  

\cameraready{At a narrower technical level, current networking progress does not seem radical. 
Ultra low-latency networks like Inifiniband are limited in scope; they require an interconnect that supports switching, which naturally incurs latency.
The limit in scope then typically translates into a need for hierarchical routing to scale horizontally, which gives heterogenity of latency. 
Meanwhile, other technologies will improve alongside networks, including direct-attached storage like HBM and NVRAM.}

\vspace{1em}
\noindent
\emph{``The main point is simple economics: Serverless is inevitable.''}
\newline\noindent
\cameraready{
Some have viewed this paper as a negative take on serverless computing, and from that perspective see the paper on the wrong side of history. 
After all, various industry-watchers have described the economic inevitability of serverless computing. To quote one such:}

\begin{quoting}[leftmargin=0.1in]
{
\it
\cameraready{I didn't have to worry about building a platform and the concept of a server, capacity planning and all that ``yak shaving'' was far from my mind... However, these changes are not really the exciting parts. The killer, the gotcha is the billing by the function...}

\cameraready{This is like manna from heaven for someone trying to build a business. Certainly I have the investment in developing the code but with application being a variable operational cost then I can make a money printing machine which grows with users...}

\cameraready{[E]xpect to see the whole world being overtaken by serverless by 2025~\cite{swardley16}.}
}
\end{quoting}

\cameraready{
It is not our intent to pour cold water on this vision. 
To reiterate, we see autoscaling (and hence pay per use) as a big step forward, but disappointingly limited to applications that can work over today's hobbled provider infrastructure. 
We acknowledge that there is an enormous market of such ``narrow'' applications, many of which consist of little more than business logic over a database.
Disrupting these applications by changing their economics will shift significant spending from traditional enterprise vendors to new, more efficient cloud-based vendors.
}

\cameraready{
However, this business motion will not accelerate the sea change in computing that the cloud offers. 
Specifically, it will not encourage---and may even deter---third-party and open source development of new stateful services, which are the core of modern computing.  Meanwhile, with innovation deterred, the cloud vendors increase market dominance for their proprietary solutions.
This line of reasoning may suggest that serverless computing could produce a local minimum: yet another setting in which the compute and storage potential of the cloud is lost in the noise of refactoring low-tech and often legacy use cases. 
}

\cameraready{
The goal of our discussion here---and, we hope, the goal of our intended audience---is to push the core technology down the playing field, rather than bet on it from the sidelines. 
To that end, we hope this paper shifts the discussion from ``What is serverless?'' or ``Will serverless win?'' to a rethinking of how we design infrastructure and programming models to spark real innovation in data-rich, cloud-scale systems. We see the future of cloud programming as far, far brighter than the promise of today's serverless FaaS offerings. Getting to that future requires revisiting the designs and limitations of what is being called ``serverless computing'' today.
}

\section{Stepping Forward to the Future}
We firmly believe that cloud programmers---whether they are writing simple applications or complex systems---need to be able to harness the compute power and storage capacity of the cloud in an autoscaling, cost-efficient manner. 
Achieving these goals requires a programmable framework that goes beyond FaaS, to dynamically manage the allocation of resources in order to meet user-specified performance goals for both compute and for data. 
 
%
Here we identify some  key challenges that remain in achieving a truly programmable environment for the cloud.

    \noindent \textbf{Fluid Code and Data Placement.} To achieve good performance, the infrastructure should be able and willing to \emph{physically colocate} certain code and data. 
    This is often best achieved by shipping code to data, rather than the current FaaS approach of pulling data to code. 
    At the same time, elasticity requires that code and data be \emph{logically separated}, to allow infrastructure to adapt placement: sometimes data needs to be replicated or repartitioned to match code needs. 
    In essence, this is the traditional challenge of data independence, but at extreme and varying scale, with multi-tenanted usage and fine-grained adaptivity in time. 
    High-level, data-centric DSLs---e.g., SQL+UDFs, MapReduce, TensorFlow---can make this more tractable, by exposing at a high level how data flows through computation. 
    The more declarative the language, the more logical separation (and optimization search space) is offered to the infrastructure.

    \noindent \textbf{Heterogeneous Hardware Support.} Cloud providers can attract a critical mass of workloads that make specialized hardware cost-effective. 
    Cloud programmers should be able to take advantage of such resources.
    Ideally, application developers could specify applications using high-level DSLs, and the cloud providers would compile those applications to the most cost-effective hardware that meets user specified SLOs. 
    Alternatively, for certain designs it may be useful to allow developers to target code to specific hardware features by specification, to foster innovative hardware/software co-design. 
    One can have philosophical debates about whether such co-design is appropriate, or whether ``hardware independence'' is a paramount concern in the cloud. 
    In either case, recognizing hardware affinity does not mean that we advocate tight binding of hardware to services; the platform should make dynamic physical decisions about allocation of code to distinguished resources, based on logical performance requirement specs either provided by programmers or extracted from code. 
    These specs can then be leveraged for the more general, heterogeneity-aware resource space-time division multiplexing~\cite{tetrisched}.

    \noindent \textbf{Long-Running, Addressable Virtual Agents.} Affinities between code, data and/or hardware tend to recur over time. 
    If the platform pays a cost to create an affinity (e.g. moving data), it should recoup that cost across multiple requests.
    This motivates the ability for programmers to establish software agents---call them functions, actors, services, etc.---that persist over time in the cloud, with known identities. 
    Such agents should be addressable with performance comparable to standard networks. 
    However, elasticity requirements dictate that these agents be virtual and dynamically remapped across physical resources. 
    Hence we need virtual alternatives to traditional operating system constructs like ``threads'' and ``ports'': nameable endpoints in the network. 
    Various ideas from the literature have provided aspects of this idea: actors~\cite{hewitt1977viewing}, tuplespaces~\cite{carriero1989linda} pub/sub~\cite{oki1994information} and DHTs~\cite{ratnasamy2001scalable} are all examples.
    Chosen solutions need to incur minimal overhead on raw network performance.

    \noindent \textbf{Disorderly programming.} As discussed above, the requirements of distributed computing and elastic resizing require changes in programming. 
    The sequential metaphor of procedural programming will not scale to the cloud. 
    Developers need languages that encourage code that works correctly in small, granular units---of both data and computation---that can be easily moved around across time and space. 
    There are examples of these patterns in the literature---particularly in asynchronous flow-based metaphors like Functional Reactive Programming~\cite{hudak1999functional} and Declarative DSLs for Networking~\cite{loo2006declarative} and Distributed Computing~\cite{bloom}.
    A particular challenge in distributed computing is to couple these programming metaphors with reasoning about semantics of distributed data consistency; earlier work offers some answers~\cite{campbellhelland,bloom,shapiro2011conflict,ameloot2013relational} but more work is needed to enable full-service applications.

    \noindent \textbf{Flexible Programming, Common IR.} Ideally, a variety of new programming languages and DSLs will be explored in this domain.
    Still, it is burdensome for each language stack to implement a full set of the relevant optimizations (e.g., fluid code \& data, disorderly constructs).
    As a result, it would be highly beneficial to develop a common internal \emph{Intermediate Representation} (IR) for cloud execution that can serve as a compilation target from many languages.
    This IR must support pluggable code from a variety of languages, as is done by UDFs in declarative languages, or functions in FaaS.


    \noindent \textbf{Service-level objectives \& guarantees}: 
    Today, none of the major FaaS offerings has APIs for service-level objectives. 
    Price is simply a function of the ``size'' (RAM, number of cores) and running time used. 
    To support practical use, FaaS offerings should enable up-front SLOs that are priced accordingly, with appropriate penalties for mis-estimation. 
    Achieving predictable SLOs requires a smooth ``cost surface'' in optimization---non-linearities are the bane of SLOs. 
    This reinforces our discussion above regarding small granules of code and data with fluid placement and disorderly execution.

    \cameraready{
    \noindent \textbf{Security concerns}: 
    Cloud programming brings up both opportunities and challenges related to security. 
    Cloud-managed software infrastructure shifts the onus of security onto a small number of well-incentivized operations staff at the cloud provider. 
    This---along with appropriate customer specifications of policy---should in principle mitigate many security issues that commonly occur today due to misconfiguration or mismanagement.
    Furthermore, if cloud programming is achieved via higher-level abstractions, it will offer the opportunity for program analysis and constraint enforcement that could improve security. 
    However, some of our desired architectural improvements for performance in this paper make achieving security more difficult for the responsible parties. 
    For example, allowing code to move fluidly toward shared data storage is potentially tricky: it exacerbates security management challenges related to multitenancy and the potential for rogue code to gather signals across customers. 
    But there are new research opportunities for innovation in this space. 
    Technologies like hardware enclaves can help protect running code, and there has been initial work on data processing in those settings (e.g.,~\cite{zheng2017opaque}). 
    Meanwhile, it is important for researchers and developers to think not only about preventative technologies, but also ways to guarantee auditing and post-hoc security analysis, as well as technologies that enable more fine-grained and easy-to-use end-user control over policy.
    }

\vspace{1em}
Taken together, these challenges seem both interesting and surmountable. 
The FaaS platforms from cloud providers are not fully open source, but the systems issues delineated above can be explored in new systems by third parties using cloud features like container orchestration. 
The program analysis and scheduling issues are likely to open up significant opportunities for more formal research, especially for data-centric programs. 
Finally, language design issues remain a fascinating challenge, bridging program analysis power to programmer productivity and design tastes.
In sum, we are optimistic that research can open the cloud's full potential to programmers. Whether we call the new results ``serverless computing'' or something else, the future is fluid.

\printbibliography

\end{document}